\documentclass{article}
\usepackage{spconf,amsmath,graphicx}
\usepackage{amssymb}
\usepackage{booktabs}
\usepackage{multirow}
\usepackage{stfloats}

\begin{document}
\ninept
\title{Fewer-token Neural Speech Codec with Time-invariant Codes}
\name{Yong Ren\textsuperscript{1,2}, Tao Wang\textsuperscript{1}, Jiangyan Yi\textsuperscript{1}, Le Xu\textsuperscript{1,2}, Jianhua Tao\textsuperscript{3}, Chu Yuan Zhang\textsuperscript{1,2}, Junzuo Zhou\textsuperscript{1,2}}
\address{\textsuperscript{1}Institute of Automation, Chinese Academy of Sciences, China \\
\textsuperscript{2}School of Artificial Intelligence, University of Chinese Academy of Sciences, China \\
\textsuperscript{3}Department of Automation, Tsinghua University, China 
}
\maketitle
\begin{abstract}
Language model based text-to-speech (TTS) models, like VALL-E, have gained attention for their outstanding in-context learning capability in zero-shot scenarios. Neural speech codec is a critical component of these models, which can convert speech into discrete token representations. However, excessive token sequences from the codec may negatively affect prediction accuracy and restrict the progression of Language model based TTS models. To address this issue, this paper proposes a novel neural speech codec with time-invariant codes named TiCodec. By encoding and quantizing time-invariant information into a separate code, TiCodec can reduce the amount of frame-level information that needs encoding, effectively decreasing the number of tokens as codes of speech. Furthermore, this paper introduces a time-invariant encoding consistency loss to enhance the consistency of time-invariant code within an utterance, which can benefit the zero-shot TTS task. Experimental results demonstrate that TiCodec can not only enhance the quality of reconstruction speech with fewer tokens but also increase the similarity and naturalness, as well as reduce the word error rate of the synthesized speech by the TTS model. The code is publicly available at https://github.com/y-ren16/TiCodec.
\end{abstract}
\section{INTRODUCTION}
\label{sec:introduction}
Recently, large language models have demonstrated remarkable performance on zero-shot text-to-speech (TTS) tasks such as VALL-E \cite{wang2023neural}, SPEAR-TTS \cite{kharitonov2023speak}, and SoundStorm \cite{borsos2023soundstorm}. 
VALL-E uses discrete tokens derived from Encodec \cite{defossez2022high} as a representation of speech, and then trains an autoregressive (AR) language model and a non-autoregressive (NAR) language model to generate tokens from the first quantizer and the other seven quantizers separately.
It can synthesize high-quality personalized speech by using a short recording of an unknown speaker as an acoustic prompt. 
However, the high-quality reconstruction of speech requires multiple frame-level token sequences, which affects the inference speed and robustness, and restricts the model structure and training methods of language model based TTS models.
Therefore, how to represent speech better with fewer tokens has become a core issue.

Neural speech codec is an important method to acquire discrete token representations of speech.
To improve the compression rate and reduce the number of tokens, more and more research is focusing on neural speech codec \cite{kumar2023high, xu23_interspeech, zheng23c_interspeech}.
Kleijn et al. \cite{kleijn2018wavenet} proposed a low-rate speech coding architecture based on the WaveNet \cite{vanwavenet} decoder.
Lyra \cite{kleijn2021generative} encodes quantized mel-spectrogram features of speech, and then decodes them with WaveGRU \cite{chung2014empirical}. 
Subsequently, end-to-end neural speech codecs have been introduced.
Grbacea et al. \cite{garbacea2019low} used the discretized latent representations proposed in VQVAE \cite{van2017neural} as conditioning for the WaveNet decoder. 
After that, SoundStream \cite{zeghidour2021soundstream}, as a fully convolutional end-to-end universal audio codec model, was proposed, extending the VQVAE vector quantizer to a residual vector quantizer. 
Following that, Encodec \cite{defossez2022high} introduced a spectrogram-only adversarial loss, a novel gradient balancer, and a small Transformer model to further improve the performance of codec. 
HifiCodec \cite{yang2023hifi} proposes a codec model that uses group-residual vector quantization to improve the reconstruction performance of audio. It can achieve good speech reconstruction performance with only four discrete token sequences, outperforming SoundStream and Encodec. 
However, the performance of codec decreases significantly when using only one or two discrete token sequences to represent speech, making it unable to reconstruct high-quality speech.

To achieve good speech reconstruction performance with only two or even one sequence of discrete frame-level tokens, we propose a neural speech codec model with time-invariant codes named TiCodec.
Some information in a speech that does not change over time is extracted by a time-invariant representation extraction module and encoded into a fixed-length code, referred to as the time-invariant code.
This operation can reduce the amount of information that needs to be encoded in frame-level codes, forcing it to be maximally informative about time-related aspects.
After obtaining the frame-level and time-invariant features, they are separately quantized as frame-level and time-invariant tokens.
When TiCodec is used for downstream TTS tasks, the time-invariant tokens can be extracted from the prompt of target speakers, which can better maintain the timbre information of target speakers.
At the same time, fewer frame-level tokens can be used to predict by the TTS model, while maintaining a low word error rate (WER) and high quality of synthesized speech.
To make the time-invariant token representations extracted from the target speech in TTS contain more global time-invariant information, we introduce the time-invariant encoding consistency loss, hoping to improve the robustness of inference in TTS and further reduce WER.

The contributions of this paper are as follows: 
\begin{itemize}
\item 
This paper proposed a neural speech codec model named TiCodec, which can separate the time-varying and time-invariant information in speech and quantize them separately. 
\item
A time-invariant encoding consistency loss was introduced to improve the consistency of the time-invariant codes.
\end{itemize}

Experimental results on speech reconstruction and zero-shot TTS task with LibriTTS datasets \cite{zen2019libritts} show that TiCodec achieved better speech reconstruction performance with fewer tokens and improved robustness, quality, and similarity of synthesized speech in the zero-shot TTS task.

\section{PROPOSED METHOD}
\label{sec:proposed}
\begin{figure*}[t]
  \centering
  \begin{minipage}[a]{0.7\linewidth}
    \centering
    \centerline{\includegraphics[width=\linewidth]{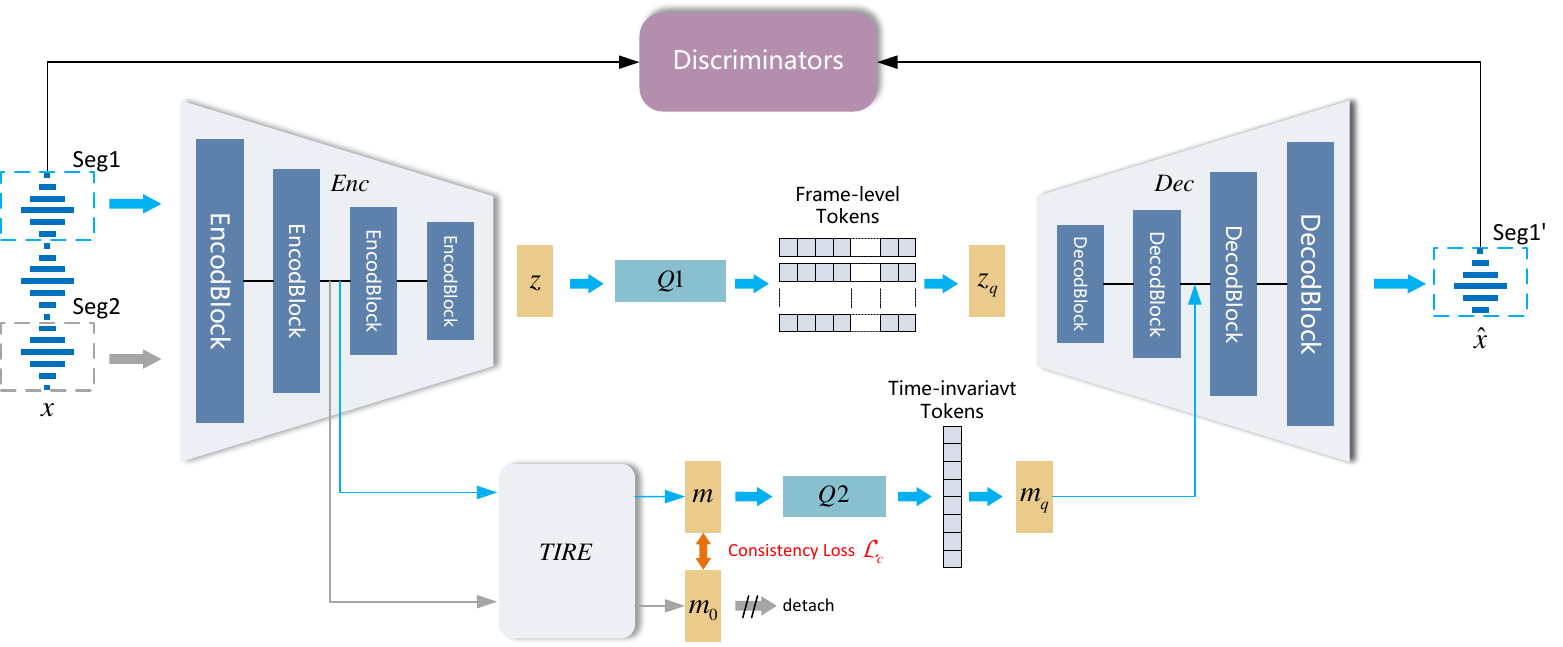}}
    \centerline{(a) The architecture of TiCodec.}\medskip
  \end{minipage}
  \begin{minipage}[a]{0.14\linewidth}
    \centering
    \vspace{0.2cm}
    \centerline{\includegraphics[width=\linewidth]{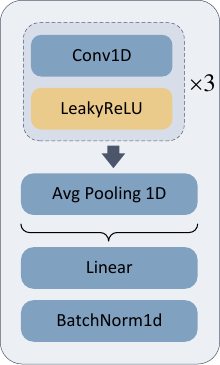}}
    \vspace{0.8cm}
    \centerline{(b) $TIRE$ module.}\medskip
  \end{minipage}
  \caption{The overview of TiCodec.}
  \label{fig:main_picture}
\end{figure*}

This paper proposes a neural speech codec model named TiCodec with a time-invariant representation extraction module and a consistency loss.
The architecture of TiCodec is shown in Figure \ref{fig:main_picture}(a).

\subsection{The Framework of TiCodec}
The overall framework of our model is an encoder-decoder architecture.
Unlike previous work, our framework utilizes a U-net-like \cite{ronneberger2015u} connection to incorporate the time-invariant features from a hidden layer of the encoder into the corresponding hidden layers of the decoder. 
We represent a speech signal of duration $d$ as $x \in \mathbb{R}^T$ with a sampling rate of $f_{sr}$, where $T=f_{sr}\times d$. 
An encoder $Enc$ transforms the input speech signal $x$ into its latent representation $z$, which is subsequently fed into a residual vector quantizer (RVQ) $Q1$ for vector quantization.
Simultaneously, the intermediate layer representation $h$ of the encoder is forwarded to the time-invariant representation extraction module ($TIRE$) to extract temporal-invariant representation $m$, which is quantized by a group vector quantizer (GVQ) $Q2$. 
The quantizer $Q1$ produces a compressed representation $z_q$ and a discrete token sequence $c_f$. And the quantizer $Q2$ produces a compressed representation $m_q$ and a discrete token sequence $c_g$.
Finally, the decoder $Dec$ reconstructs the speech signal $\hat{x}$ from the compressed latent representation $z_q$ and the compressed time-invariant representation $m_q$. 

$Enc$ and $Dec$ follow a similar structure as HifiCodec \cite{yang2023hifi}. 
$Enc$ is composed of a 1D convolutional layer followed by 4 convolutional modules and a final 1D convolutional layer. 
Each convolutional module consists of three residual units and one downsampling layer. 
All of these 4 modules indicate a total downsampling of 320 times.
$Dec$ adopts a symmetric structure to the encoder, utilizing transpose convolutions for upsampling.
$Q1$ is a RVQ module to quantize the latent representation of the speech to frame-level tokens.
We used the same discriminators as HifiCodec \cite{yang2023hifi}, which include three discriminators: a multi-scale STFT-based discriminator, a multi-period discriminator, and a multi-scale discriminator \cite{kong2020hifi}.

\subsection{Time-invariant Representation Extraction and Quantization}
\label{ssec:ticem}
We use the output of the second convolutional module in $Enc$ as the input to $TIRE$ to extract time-invariant representation $m$. 
Similarly, the compressed time-invariant representation $m_q$ after quantizer $Q2$ quantized is introduced into the second transpose convolutional module of $Dec$ for decoding. 

As shown in Figure \ref{fig:main_picture}(b), $TIRE$ first performs further feature extraction on the input speech features through three 1D convolutional layers and LeakyReLU layers.
Next, temporal averaging on the extracted features was employed to summarize the frame-level features into a time-invariant feature.
Then, we pass this representation through a fully connected layer and an activation function to obtain the final time-invariant representation $m$.

For the quantization of the time-invariant representation $m$, we use GVQ, which divides $m$ into eight groups, resulting in eight discrete tokens as the time-invariant codes.
GVQ expands the representation space, enabling a more extensive portrayal of time-invariant encoding.

Then the compressed time-invariant representation $m_q$ is duplicated across the temporal dimension, converting segment-level time-invariant features back to frame-level features.
Afterward, they are added to the input of the penultimate layer in the original decoder to perform decoding jointly.

The time-invariant representation extraction and quantization modules establish a connection between the contracting and expansive paths, enabling the flow of shallow representations from the contracting path to the symmetric expansive path through the extraction of time-invariant features.
Simultaneously representing time-invariant information through time-invariant codes, the remaining frame-level codes tend to capture greater temporal dependencies as a result of the information bottleneck, thereby reducing redundancy frame-level codes.
\subsection{Time-invariant Encoding Consistency Loss}
\label{sssec:cosloss}
When TiCodec is employed for downstream TTS tasks, We extract time-invariant tokens from the target speech segment and then use textual information to predict frame-level tokens.
In order to maintain consistency of invariant codes extracted from different segments of an utterance, we propose the time-invariant encoding consistency Loss ($\mathcal{L}_c$). 
During training, in addition to encoding the speech segment $seg_1$, we randomly sample another segment $seg_2$ from the same utterance. 
Then, $seg_2$ also goes through the first two convolution modules of $Enc$ and $TIRE$, followed by a stop-gradient operation. 
The type of training is shown in Figure \ref{fig:main_picture}(a).

We use cosine similarity loss as the consistency loss for the extracted time-invariant representations of the two segments, denoted as $\mathcal{L}_c$.
\begin{equation}
  \begin{aligned}
    \mathcal{L}_c=1-cos(TIRE(Enc[:2](x_1)),TIRE(Enc[:2](x_2)))
  \end{aligned}
\end{equation}
where $x_1$ and $x_2$ denote $seg_1$ and $seg_2$ of speech waveform separately and $cos$ denotes the cosine similarity function.

The generator is trained to optimize the following loss:
\begin{equation}
\begin{split}
  \mathcal{L} & =\lambda_t \mathcal{L}_t + \lambda_f \mathcal{L}_f + \lambda_g \mathcal{L}_g + \lambda_{feat} \mathcal{L}_{feat} \\
  & + \lambda_{qz} \mathcal{L}_{qz} + \lambda_{qm} \mathcal{L}_{qm} + \lambda_{c} \mathcal{L}_c
\end{split}
\end{equation}
where $\mathcal{L}_t$ and $\mathcal{L}_f$ are reconstruction losses of time domain and frequency domain, 
$\mathcal{L}_g$ is the adversarial loss of the generator, 
$\mathcal{L}_{feat}$ is the feature matching loss, and
$\mathcal{L}_{qz}$ is the quantization loss of frame-level codes.
These losses are the same as that in HifiCodec \cite{yang2023hifi}.
$\mathcal{L}_{qm}=||m-m_{q}||^2$ is the commitment loss of time-invariant codes, and
$\mathcal{L}_c$ is the time-invariant encoding consistency loss.
$\lambda_t$, $\lambda_f$, $\lambda_g$, $\lambda_{feat}$, $\lambda_{qz}$, $\lambda_{qm}$ and $\lambda_{c}$ are hyperparameters to balance each term of the final loss.

\section{EXPERIMENTS}
\label{sec:experiments}

\subsection{EXPERIMENTS SETUP}
\label{ssec:setup}
\textbf{Dataset.}
We used LibriTTS \cite{zen2019libritts} datasets to train TiCodec we proposed and VALL-E model.
The LibriTTS corpus consists of 585 hours of speech data at 24kHz from 2,456 speakers. 
Our training set was a combination of the train-clean-100, train-clean-360, and train-other-500 subsets. 
To evaluate our model, we randomly sampled 100 utterances from test subsets of LibriTTS, VCTK \cite{liu2019cross} and AISHELL3 \cite{shi2020aishell} datasets separately.

\noindent
\textbf{Baselines.}
We utilized Encodec \cite{defossez2022high} and HifiCodec \cite{yang2023hifi} as the baseline for our TiCodec model. For TTS model VALL-E, we used HifiCodec \cite{yang2023hifi} as the baseline to convert speech to discrete token representations.

\noindent
\textbf{Training.} The downsampling factors of the encoder of codecs were set as [8,5,4,2], resulting in a total downsampling of 320 times. 
The batch size was 40, and the learning rate was 0.002.
Our model and HifiCodec were trained for 250k iterations when using 2 and 4 frame-level quantizers, and 300k iterations when using 1 frame-level quantizer.
The baseline model Encodec was used with the code reimplemented by Yang et al. \cite{yang2023hifi} in HifiCodec\footnote[1]{https://github.com/yangdongchao/AcademiCodec}, and it was trained for 25 epochs.
For VALL-E\footnote[2]{https://github.com/lifeiteng/vall-e}, we set the maximum duration per batch to 160. 
The AR stage was trained for 30 epochs with a maximum learning rate of 0.05.
The NAR stage was trained for 40 epochs with a maximum learning rate of 0.01.
We trained all models on a single A100 GPU.

\noindent
\textbf{Evaluation Metrics.}
ViSQOL V3 \cite{chinen2020visqol}, PESQ \cite{recommendation2001perceptual}, STOI \cite{taal2010short}, and Mel cepstral distortion (MCD) are used to evaluate the objective quality of speech reconstructed by TiCodec and baselines.

We consider both objective and subjective metrics for VALL-E.
We use Resemblyzer\footnote[3]{https://github.com/resemble-ai/Resemblyzer} to derive a high-level representation of a voice and then evaluate the speaker similarity between the prompt and synthesized speech.
The content accuracy was evaluated by transcribing the speech with hubert-large-ls960-ft \footnote[4]{https://huggingface.co/facebook/hubert-large-ls960-ft} \cite{hsu2021hubert} model and calculating the WER.
Additionally, we calculate the Mean Opinion Score (MOS) and Similarity Mean Opinion Score (SMOS) with 95\% confidence by crowdsourcing. 8 participants were invited to test speech naturalness and speaker similarity, with MOS and SMOS results.
\subsection{Results}
\label{ssec:result}

\subsubsection{Speech Reconstruction Performance of Codecs}
\label{sssec:codec}

\begin{table}
  \centering
  \setlength{\tabcolsep}{1.2mm}{
    \caption{Objective metrics scores of various codecs on LibriTTS.}
  \begin{tabular}{clcccc}
  \toprule
  $n_q$ & Model & ViSQOL $\uparrow$ & PESQ $\uparrow$ & STOI $\uparrow$ & MCD $\downarrow$ \\
    \midrule
    \multirow{4}{*}{1} & Encodec & 2.445 & 1.202 & 0.749 & 1.490 \\
      & HifiCodec & 3.335 & 1.454  & 0.817 & 1.416 \\
      & TiCodec (Ours) & \bf{3.631} & \bf{1.663}  & \bf{0.855} & 1.315 \\
      & TiCodec+$\mathcal{L}_c$ (Ours) & 3.615 & 1.634  & 0.854 & \bf{1.312} \\
    \midrule
    \multirow{4}{*}{2} & Encodec & 2.562 & 1.272 & 0.783 & 1.356 \\
      & HifiCodec & 3.824 & 2.025  & 0.888 & 1.055 \\
      & TiCodec (Ours) & \bf{3.944} & \bf{2.149}  & \bf{0.903} & 0.985 \\
      & TiCodec+$\mathcal{L}_c$ (Ours) & 3.926 & 2.124  & \bf{0.903} & \bf{0.981} \\
    \midrule
    \multirow{4}{*}{4} & Encodec & 2.579 & 1.290 & 0.789 & 1.330 \\
      & HifiCodec & 4.243 & 2.697  & 0.936 & 0.758 \\
      & TiCodec (Ours) & 4.244 & \bf{2.773}  & 0.940 & 0.737 \\
      & TiCodec+$\mathcal{L}_c$ (Ours) & \bf{4.255} & 2.734  & \bf{0.941} & \bf{0.735} \\
    \bottomrule
  \end{tabular}
  \label{tbl:table1}
  }
\end{table}

To evaluate the speech reconstruction quality of TiCodec, we compared it with the baseline Encodec and HifiCodec with 1, 2, and 4 quantizers separately. Number of frame-level quantizers is denoted as $n_q$.
The objective scores evaluated on a test set of LibriTTS are shown in Table \ref{tbl:table1}.

Table \ref{tbl:table1} shows that TiCodec outperforms the baseline in all objective metrics. 
Furthermore, as the number of quantized tokens decreases, the performance improvement of TiCodec becomes increasingly significant.
This result is reasonable because a decrease in the number of frame-level quantizers results in a limited number of codes becoming increasingly inadequate for encoding the entirety of the information of the original speech.
The introduction of time-invariant codes eliminates the need for duplicating encoding of repetitive information in speech, such as timbre and acoustic environment, at the frame level. This improves the coding efficiency.

To verify the generalization of TiCodec, we also evaluated the performance of different codecs on the VCTK and AISHELL3 datasets. 
The objective scores are shown in Table \ref{tbl:table2}. 
We can see that our model also outperforms the baseline model in all objective metrics on both datasets. This indicates that our model is quite competitive in generalization ability.
\begin{figure}[h]
  \centering
  \begin{minipage}[t]{0.4\linewidth}
    \begin{minipage}[t]{0.3\linewidth}
      \centering
      \centerline{\includegraphics[width=\linewidth]{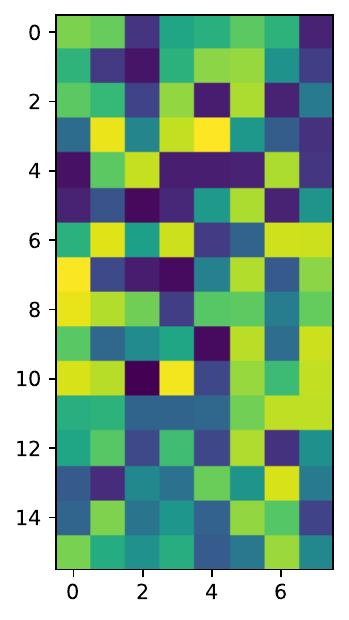}}
      \centerline{Full}\medskip
    \end{minipage}
    \begin{minipage}[t]{0.3\linewidth}
      \centering
      \centerline{\includegraphics[width=\linewidth]{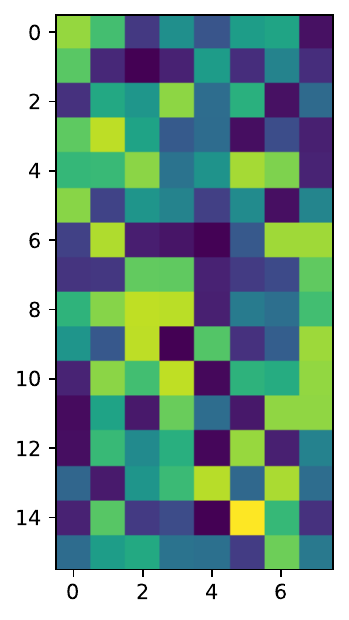}}
      \centerline{Seg-F}\medskip
    \end{minipage}
    \begin{minipage}[t]{0.3\linewidth}
      \centering
      \centerline{\includegraphics[width=\linewidth]{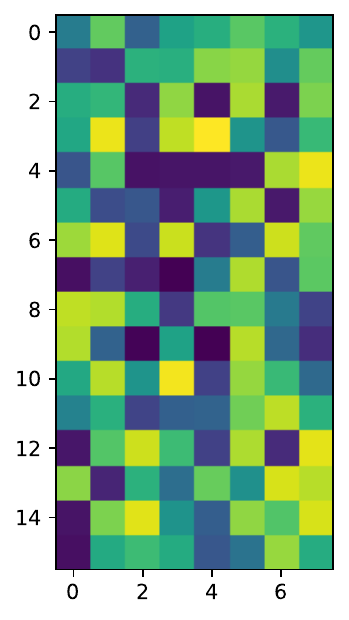}}
      \centerline{Seg-L}\medskip
    \end{minipage}
    \centerline{(a) TiCodec}\medskip
  \end{minipage}
  \hspace{5mm}
  \begin{minipage}[t]{0.4\linewidth}
    \begin{minipage}[t]{0.3\linewidth}
      \centering
      \centerline{\includegraphics[width=\linewidth]{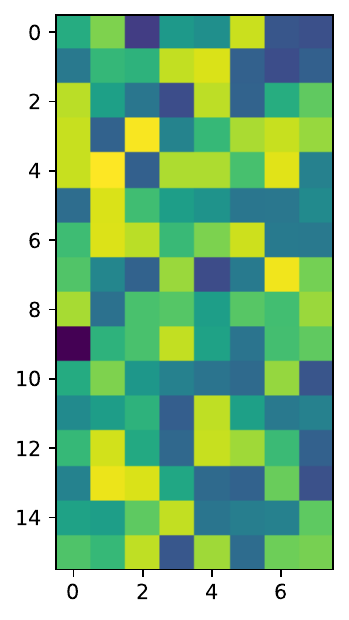}}
      \centerline{Full}\medskip
    \end{minipage}
    \begin{minipage}[t]{0.3\linewidth}
      \centering
      \centerline{\includegraphics[width=\linewidth]{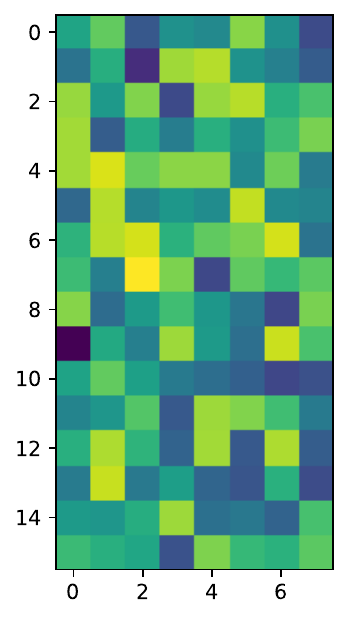}}
      \centerline{Seg-F}\medskip
    \end{minipage}
    \begin{minipage}[t]{0.3\linewidth}
      \centering
      \centerline{\includegraphics[width=\linewidth]{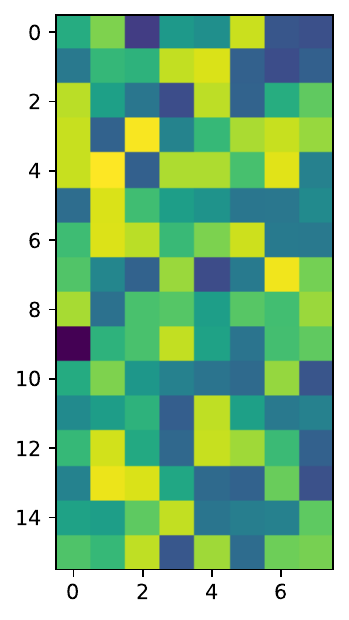}}
      \centerline{Seg-L}\medskip
    \end{minipage}
    \centerline{(b) TiCodec+$\mathcal{L}_c$}\medskip
  \end{minipage}
  \caption{The time-invariant codes extracted from an utterance.}
  \label{fig:frature_picture}
\end{figure}

We can see from Table \ref{tbl:table1} and Table \ref{tbl:table2} that, after introducing the time-invariance encoding consistency loss $\mathcal{L}_c$, the performance of the TiCodec slightly decreases on ViSQOL and PESQ metrics. It indicates that $\mathcal{L}_c$ slightly affects the perceived quality of the reconstructed speech.
Figure \ref{fig:frature_picture} presents the visualization of the time-invariant representations $m$ extracted from an utterance in the test set.
The three representations in Figure \ref{fig:frature_picture}(a) and Figure\ref{fig:frature_picture}(b) were extracted from the entire, the first half (Seg-F), and the last half (Seg-L) of the utterance. 
Representations in Figure \ref{fig:frature_picture}(b) and Figure\ref{fig:frature_picture}(a) were extracted from TiCodec with and without $\mathcal{L}_c$, respectively. 
By adding $\mathcal{L}_c$, the consistency between representations of different segments within the same utterance is improved. This is advantageous for the zero-shot TTS task, as demonstrated in the following section.

\begin{table*}[t]
  \centering
  \setlength{\tabcolsep}{1mm}{
  \caption{The objective metrics scores testing on AISHELL3 and VCTK datasets of various codecs trained on LibriTTS.}
  \begin{tabular}{clcccc|cccc}
  \toprule
  \multirow{2}{*}{$n_q$} & \multirow{2}{*}{Model} & \multicolumn{4}{c}{AISHELL3} & \multicolumn{4}{c}{VCTK} \\
  & & ViSQOL $\uparrow$ & PESQ $\uparrow$ & STOI $\uparrow$ & MCD $\downarrow$ & ViSQOL $\uparrow$ & PESQ $\uparrow$ & STOI $\uparrow$ & MCD $\downarrow$\\
    \midrule
    \multirow{4}{*}{1}& Encodec & 2.175 & 1.124  & 0.651  & 1.007 & 2.565 & 1.251  & 0.688 & 1.240 \\
      & HifiCodec & 3.005 & 1.319  & 0.732  & 0.955 & 3.267 & 1.427  & 0.748 & 1.120 \\
      & TiCodec (Ours) & \bf{3.390} & \bf{1.434}  & \bf{0.777}  & 0.902 & \bf{3.613} & 1.641  & 0.795 & 0.980 \\
      & TiCodec+$\mathcal{L}_c$ (Ours) & 3.377 & 1.424  & 0.775  & \bf{0.855} & 3.588 & \bf{1.723}  & \bf{0.801} & \bf{0.978} \\
    \cline{1-10}
    \multirow{4}{*}{2} & Encodec & 2.325 & 1.165  & 0.692  & 0.934 & 2.682 & 1.305  & 0.710 & 1.193 \\
      & HifiCodec & 3.625 & 1.677  & 0.820  & 0.733 & 3.691 & 1.721  & 0.811 & 0.891 \\
      & TiCodec (Ours) & \bf{3.743} & \bf{1.750}  & 0.836 & 0.673 & \bf{3.837} & \bf{1.848}  & \bf{0.831} & \bf{0.824} \\
      & TiCodec+$\mathcal{L}_c$ (Ours) & 3.738 & 1.738  & \bf{0.841}  & \bf{0.668} & 3.802 & 1.805  & 0.826 & 0.863 \\
    \cline{1-10}
    \multirow{4}{*}{4} & Encodec & 2.349 & 1.173  & 0.698  & 0.917 & 2.695 & 1.310  & 0.714 & 1.180 \\
      & HifiCodec & 4.108 & 2.154  & 0.892  & 0.502 & 4.134 & 2.091  & 0.868 & 0.705 \\
      & TiCodec (Ours) & 4.114 & \bf{2.221}  & 0.894  & 0.507 & 4.150 & \bf{2.266}  & 0.872 & 0.676 \\
      & TiCodec+$\mathcal{L}_c$ (Ours) & \bf{4.122} & 2.162  & \bf{0.896} & \bf{0.490} & \bf{4.155} & 2.177  & \bf{0.878} & \bf{0.650} \\
    \bottomrule
  \end{tabular}
  \label{tbl:table2}
  }
\end{table*}

\subsubsection{Downstream zero-shot TTS task}
\label{sssec:downstream}
\begin{figure}[t]
  \centering
  \centerline{\includegraphics[width=\linewidth]{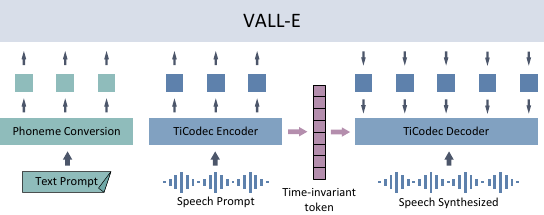}}
  \caption{VALL-E used TiCodec as tokenizer.}
  \label{fig: VALL-E used TiCodec}
\end{figure}

\begin{table}[t]
  \centering
  \setlength{\tabcolsep}{1.2mm}{
  \caption{The objective and subjective metrics scores of VALL-E.}
  \resizebox{!}{0.3\linewidth}{
  \begin{tabular}{clcccc}
  \toprule
  \multirow{2}{*}{$n_q$} & \multirow{2}{*}{Codec Model} & \multicolumn{2}{c}{Objective} & \multicolumn{2}{c}{Subjective} \\
  & & WER $\downarrow$ & SIM $\uparrow$  & MOS $\uparrow$ & SMOS $\uparrow$\\
    \midrule
    \multicolumn{2}{c}{Groundtruth} & 4.60 & - & 4.50 & - \\
    \midrule
    \multirow{3}{*}{1} & HifiCodec & 55.5 & 0.701 & 3.28$_{(\pm 0.20)}$ & 3.08$_{(\pm 0.14)}$ \\
    & TiCodec (Ours) & 34.8 & \bf{0.755} & 3.80$_{(\pm 0.21)}$ & 4.13$_{(\pm 0.14)}$ \\
    & TiCodec+$\mathcal{L}_c$ (Ours)  & \bf{28.8} & 0.7470 & \bf{4.10}$_{(\pm \bf{0.20})}$ & \bf{4.23}$_{(\pm \bf{0.12})}$ \\
    \midrule
    \multirow{3}{*}{2} & HifiCodec & 18.8 & 0.736 & 3.95$_{(\pm 0.19)}$ & 3.73$_{(\pm 0.15)}$ \\
    & TiCodec (Ours) & 12.4 & \bf{0.770} & \bf{4.40}$_{(\pm \bf{0.16})}$ & \bf{4.15}$_{(\pm \bf{0.12})}$ \\
    & TiCodec+$\mathcal{L}_c$ (Ours) & \bf{9.10} & 0.759 & 4.15$_{(\pm 0.17)}$ & 3.95$_{(\pm 0.14)}$ \\
    \midrule
    \multirow{3}{*}{4} & HifiCodec & 12.0 & \bf{0.7917} & 4.08$_{(\pm 0.17)}$ & 4.18$_{(\pm 0.09)}$ \\
    & TiCodec (Ours) & 12.6 & 0.780 & \bf{4.18}$_{(\pm \bf{0.18})}$ & 4.23$_{(\pm 0.11)}$ \\
    & TiCodec+$\mathcal{L}_c$ (Ours)  & \bf{11.2} & 0.782 & 4.00$_{(\pm 0.19)}$ & \bf{4.28}$_{(\pm \bf{0.11})}$ \\
    \bottomrule
  \end{tabular}
  }
  \label{tbl:table3}
  }
\end{table}	


We trained VALL-E using the discrete tokens of speech extracted separately from HifiCodec, TiCodec, and TiCodec with $\mathcal{L}_c$.
For TiCodec, we directly extract the time-invariant tokens from the prompt speech and incorporate them into the decoder, as shown in Figure \ref{fig: VALL-E used TiCodec}.

Table \ref{tbl:table3} displays the performance of different neural codecs used for VALL-E. 
TiCodec demonstrates a lower WER and higher speaker similarity than HifiCodec when used as a tokenizer for VALL-E.
The MOS and SMOS score of VALL-E with TiCodec is significantly higher than that of HifiCodec, especially when the number of frame-level quantizers is 1 and 2.
The incorporation of the time-invariant encoding consistency loss further reduces the WER.
This improvement occurs because time-invariant tokens are extracted from prompt audio, and frame-level time-varying tokens are predicted.
Consequently, the consistency loss enhances the uniformity of time-invariant tokens, thereby improving inference stability.
We put reconstructed speech from various codec models and samples of zero-shot TTS on our demo page. \footnote[5]{https://y-ren16.github.io/TiCodec}  

\section{Conclusion}
\label{sec:conclusion}
In this study, we proposed a novel neural speech codec model called TiCodec with a time-invariant representation extraction module and a time-invariant encoding consistency loss.
Our model exhibits improved performance in speech reconstruction using fewer tokens, surpassing the performance of previous codecs.
By utilizing the extracted tokens from TiCodec as the input of the zero-shot TTS model VALL-E,
we achieved a reduction in WER and enhanced the naturalness and similarity of the synthesized speech.
In the future, we will explore better methods of decoupling temporal information from invariant information and more effective zero-shot TTS methods to utilize fewer tokens extracted by TiCodec.

\section{ACKNOWLEDGMENTS}
\label{sec:acknowledgments}
This work is supported by the Scientific and Technological Innovation Important Plan of China (No. 2021ZD0201502), the National Natural Science Foundation of China (NSFC) (No. 62322120, No. 62306316, No.61831022, No.U21B2010, No.62101553, No.61971419, No.62006223, No. 62206278).
\vfill\pagebreak

\bibliographystyle{IEEEbib}
\bibliography{refs}

\end{document}